# Near-field radiative heat transfer in the dual nanoscale regime between polaritonic membranes

Lívia Corrêa McCormack[1], Lei Tang[2] and Mathieu Francoeur[1*]

[1]Department of Mechanical Engineering, The University of Utah, Salt Lake City, UT 84112, USA

[2]Department of Mechanical Engineering, University of California at Berkeley, Berkeley, CA 94720, USA

**ABSTRACT**

The enhancement and attenuation of near-field radiative heat transfer between polaritonic SiC, SiN and $SiO_2$ subwavelength membranes is analyzed. Fluctuational electrodynamics simulations combined with a modal analysis show that all membranes support corner and edge modes, which can induce a large 5.1-fold enhancement for SiC and a 2.1-fold attenuation for $SiO_2$ of the heat transfer coefficient with respect to that between infinite surfaces. The enhancement or attenuation is directly related to material losses which reduce the density of available electromagnetic states between the membranes.

[*]Corresponding author: mfrancoeur@mech.utah.edu

According to Planck's theory of thermal radiation, the blackbody concept is valid when all system length scales are much larger than the thermal photon wavelength $\lambda_{th}$ given by Wien's law [1]. When the gap spacing $d$ between the thermal sources is smaller than $\lambda_{th}$, it is well-established both theoretically [2-5] and experimentally [6-14] that near-field radiative heat transfer (NFRHT) can exceed the blackbody limit owing to the tunneling of evanescent electromagnetic waves. In the far-field regime ($d \gg \lambda_{th}$) where radiative heat transfer is solely mediated by propagating electromagnetic waves, experiments [15] and theory [16] have also demonstrated that the blackbody limit can be surpassed between membranes thinner than $\lambda_{th}$. This is because the absorption cross section $C_{abs}$ of an object with characteristic dimension smaller than $\lambda_{th}$ can exceed its geometrical cross section $A_c$, which results in an enhancement beyond the blackbody limit based on $A_c$. This phenomenon has long been recognized in the light scattering community where the emissivity of subwavelength particles with $C_{abs} > A_c$ exceeds unity [17,18]. A few other works have analyzed far-field radiative heat transfer between rectangular [19-22] and conical [23] subwavelength membranes.

Recently, NFRHT in the dual nanoscale regime between two subwavelength membranes separated by a subwavelength vacuum gap has been studied [24,25]. Tang et al. [24] reported NFRHT between two SiC membranes with thickness approximately equal to or smaller than the gap spacing of 100 nm. For 20-nm-thick membranes, the heat transfer coefficient was measured and predicted to be ~5.5 times larger than that between two infinite surfaces separated by the same gap, and ~1400 times larger than the blackbody limit accounting for the view factor. The enhancement was ascribed to evanescent electromagnetic corner and edge modes [26] generated by the coupling of surface phonon-polaritons (SPhPs) in the membranes. Luo at al. [25] measured and predicted NFRHT between two 300-nm-thick SiN membranes, a material that also supports

SPhPs in the infrared, separated by gaps ranging from 150 to 750 nm. They observed a 20-fold enhancement of the flux beyond the blackbody limit accounting for the view factor. However, the radiative transfer between the membranes was lower than that between two infinite surfaces separated by the same gap.

On one hand, SiC and SiN are both polaritonic materials exhibiting similar electromagnetic responses. On the other hand, in the dual nanoscale regime, SiC and SiN membranes display seemingly opposite trends in NFRHT. This raises the following scientific question: *Why does the NFRHT coefficient between polaritonic subwavelength membranes not always surpass that between two infinite surfaces?* The objective of this Letter is to answer this question by analyzing the physics of NFRHT in the dual nanoscale regime between polaritonic subwavelength membranes made of SiC, SiN and $SiO_2$. By combining numerically-exact discrete system Green's function (DSGF) simulations of NFRHT [27,28] with a modal analysis, it is shown that all membranes support electromagnetic corner and edge modes. Depending on material losses that reduce the available density of electromagnetic states, these modes can lead to an enhancement or attenuation of the heat transfer coefficient with respect to that between two infinite surfaces. This study can impact future thermal photonic-based technologies, such as localized radiative cooling [29] and solid-state energy conversion devices [30,31].

The system under study, shown in Fig. 1, consists of two coplanar subwavelength membranes separated by a fixed vacuum gap $d = 100$ nm. The membranes' width $w$ and length $L$ are fixed at 1 µm, whereas their thickness $t$ varies from 1000 nm ($t >> d$) down to 20 nm ($t << d$). NFRHT is calculated using the fluctuational electrodynamics-based [32] DSGF method [27] in which the membranes are discretized into cubic subvolumes of size $\Delta V_i$ smaller than the vacuum and material wavelengths. In that way, the electric field and Green's functions can be approximated to be uniform in $\Delta V_i$. The total thermal conductance between the hot-emitting membrane at temperature $T + \delta T$ and the cold-receiving membranes at temperature $T$ in the limit that $\delta T \to 0$ is calculated as follows:

$$G_{\text{rad}}(T) = \frac{1}{2\pi} \int_0^\infty \left.\frac{\partial \Theta(\omega, T')}{\partial T}\right|_{T'=T} \mathcal{T}(\omega) d\omega, \tag{1}$$

where $\Theta(\omega, T)$ is the mean energy of an electromagnetic state. The spectral transmission coefficient between the hot and cold membranes of respective volumes $V_h$ and $V_c$ is given by:

$$\mathcal{T}(\omega) = \sum_{i \in V_h} \sum_{j \in V_c} 4k_0^4 \Delta V_i \Delta V_j \text{Im}[\varepsilon(\mathbf{r}_i, \omega)] \text{Im}[\varepsilon(\mathbf{r}_j, \omega)] \text{Tr}[\overline{\overline{\mathbf{G}}}(\mathbf{r}_i, \mathbf{r}_j, \omega) \overline{\overline{\mathbf{G}}}^\dagger(\mathbf{r}_i, \mathbf{r}_j, \omega)] d\omega, \tag{2}$$

where $k_0$ is the vacuum wave vector magnitude, $\varepsilon$ is the dielectric function, † is the conjugate transpose operator, and $\overline{\overline{\mathbf{G}}}(\mathbf{r}_i, \mathbf{r}_j, \omega)$ denotes the monochromatic system Green's function relating subvolumes $i$ and $j$ evaluated according to the procedure outlined in Appendix A. The heat transfer coefficient is calculated from the thermal conductance as $h_{\text{rad}}(T) = G_{\text{rad}}(T)/A_c$, where $A_c = Lt$ is the membrane cross-sectional area. The dielectric functions of SiC, SiN, and SiO2 are provided in Sec. S1 of the Supplemental Material [33].

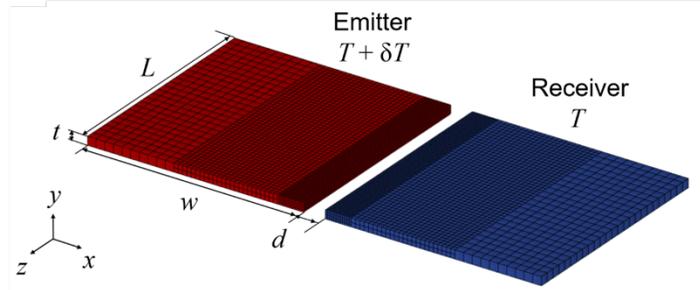

FIG. 1. NFRHT between a hot emitting membrane at temperature $T + \delta T$ and a cold receiving membrane at temperature $T$ separated by a fixed vacuum gap $d = 100$ nm. The membranes of variable thickness $t$ and fixed dimensions $w = 1$ μm and $L = 1$ μm are discretized into nonuniform subvolumes for computational efficiency.

The total heat transfer coefficient and total thermal conductance at 300 K are reported as a function of the membrane thickness in Figs. 2(a) and 2(b), respectively. The heat transfer coefficient between two infinite surfaces separated by a gap $d = 100$ nm shown in Fig. 2(a) is calculated using a closed-form expression derived from fluctuational electrodynamics (see Appendix B) [34]. Note that the DSGF convergence analysis is provided in Sec. S2 of the Supplemental Material [33].

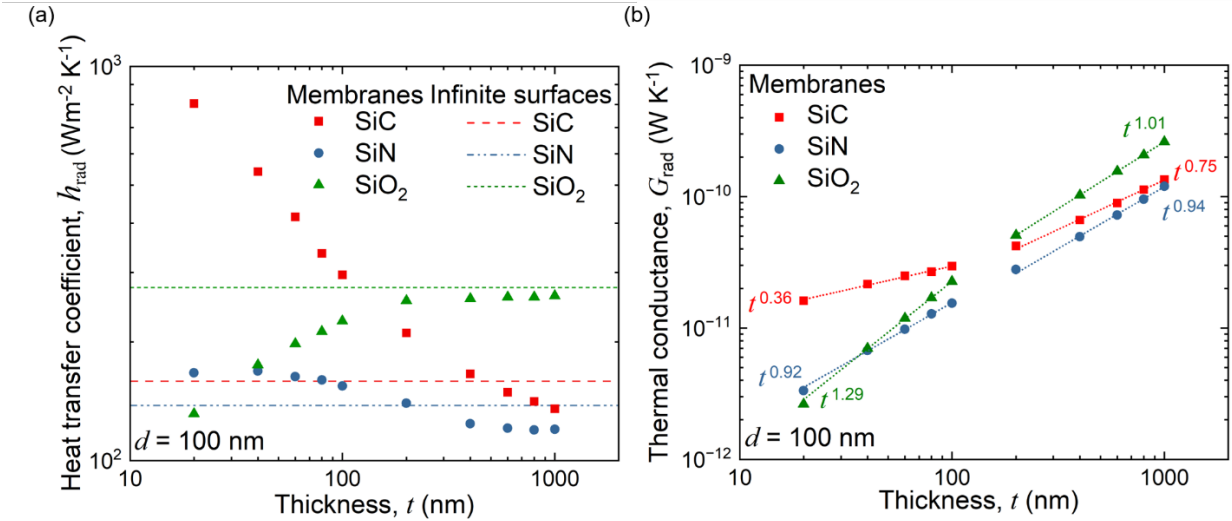

FIG. 2. NFRHT between SiC, SiN, and SiO$_2$ membranes separated by a vacuum gap $d = 100$ nm calculated with the DSGF method. (a) Heat transfer coefficient, $h_{rad}$, at 300 K as a function of the membrane thickness. Results are compared against those obtained between two infinite surfaces separated by $d = 100$ nm. (b) Thermal conductance, $G_{rad}$, at 300 K as a function of the membrane thickness. The variations of $G_{rad}$ with respect to $t$ are identified by dashed lines.

Figure 2(a) reveals different NFRHT trends in the dual nanoscale regime despite that all materials support SPhPs in the infrared. The heat transfer coefficient $h_{rad}$ between SiC membranes increases as the membrane thickness decreases, as previously measured and predicted in Ref. [24], and exceeds that between infinite surfaces for membrane thickness to gap spacing ratio, $t/d$, approximately equal to or smaller than 4 ($t \lesssim 400$ nm). Here, a maximum 5.1-fold enhancement is predicted for 20-nm-thick membranes. An $h_{\text{rad}}$ enhancement, albeit much smaller, is also observed for SiN membranes when $t/d$ is less than 2 ($t < 200$ nm). The heat transfer coefficient reaches a maximum for membrane thickness $t$ of 40 nm, and slightly diminishes when $t$ is decreased to 20 nm. This enhancement between SiN membranes was not observed in Ref. [25], which is likely due to the fact that $t/d$ was ~2 for the smallest gap spacing of 150 nm. Conversely to SiC and SiN, the heat transfer coefficient between $SiO_2$ membranes is substantially attenuated as $t$ decreases, and reaches a minimum value 2.1 times smaller than that between infinite surfaces when $t = 20$ nm. Note that $h_{\text{rad}}$ does not converge to the infinite surface cases as $t$ increases because the selected $L$ and $w$ dimensions are not large enough to recover the full electromagnetic states that exist in the infinite surface cases.

The thermal conductance $G_{\text{rad}}$ in Fig. 2(b) follow different power laws for thin ($t/d < 2$) and thick ($t/d \gtrsim 2$) membranes. For thick $SiO_2$ membranes, the thermal conductance varies linearly with thickness ($t^{1.01}$), such that $h_{\text{rad}}$, which is proportional to $G_{\text{rad}}/t$, is approximately independent of $t$. This implies that as for the case of infinite surfaces, NFRHT is dominated by the tunneling of SPhPs supported by solitary $SiO_2$-vacuum interfaces adjacent to the gap spacing and parallel to the $y$-$z$ plane. The contribution of these uncoupled SPhPs simply decreases proportionally with $t$ for a fixed $L$ value. A similar behavior is observed for thick SiN membranes for which the thermal conductance varies as $t^{0.94}$, although a small enhancement of the heat transfer is seen in Fig. 2(a) when $t$ decreases from 400 nm to 200 nm.

For thin $SiO_2$ and SiN membranes, and thick and thin SiC membranes, the dependence of thermal conductance on membrane thickness deviates appreciably from the linear regime. This suggests that NFRHT is dominated by coupled SPhP modes for those cases. Berini [26] showed that the coupling of the evanescent electromagnetic fields produced by surface plasmon-polaritons traveling along the *z*-direction (see Fig. 1) in a solitary metallic membrane of infinite length *L* generate four fundamental electromagnetic corner and edge modes. Specifically, this coupling arises between the membrane's perpendicular edges through the corners, as well as between neighboring corners. Tang et al. [24] demonstrated that these corner and edge modes are also generated in SiC membranes via the coupling of SPhPs.

The different trends observed in Fig. 2 can be explained via a modal analysis of solitary membranes of infinite length *L*. The analysis is performed by assuming a real frequency $\omega$ and a complex wave vector $k_z$. Figure 3 shows corner and edge mode dispersion relations for two fundamental modes, *aa* and *sa*, where *a* means asymmetric and *s* symmetric. For instance, the *sa* mode indicates that the *y*-component of the electric field is symmetric with respect *y* but asymmetric with respect to the *x*-axis [26]. Note that the high-frequency fundamental *ss* and *as* modes are not plotted in Fig. 3 because they do not impact significantly the NFRHT.

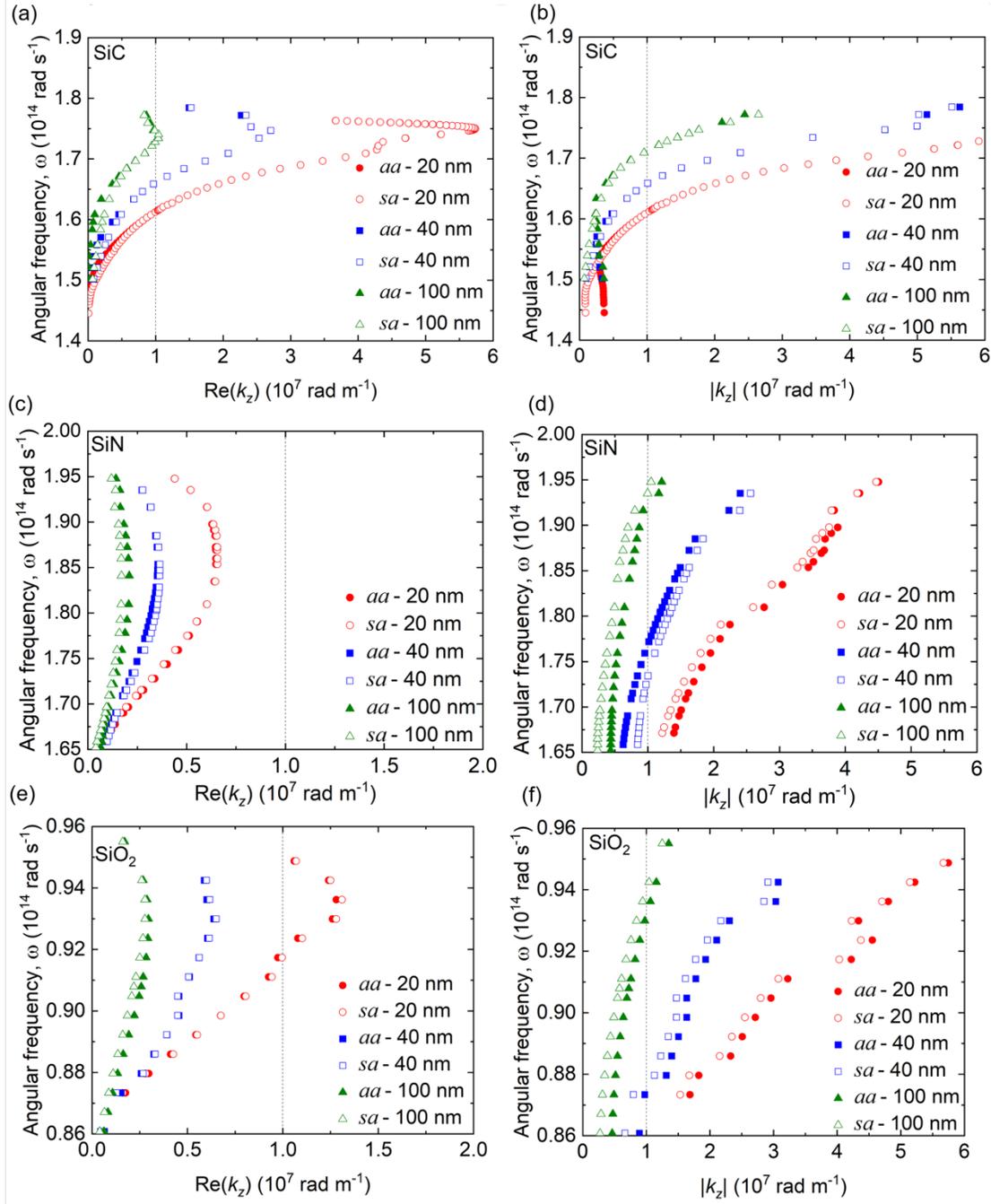

FIG 3. Dispersion relations for the two low-frequency fundamental corner and edge modes, *aa* and *sa*, in a solitary membrane of infinite length $L$ and varying thickness $t$ calculated via COMSOL Multiphysics: SiC [panels (a) and (b)], SiN [panels (c) and (d)], and SiO$_2$ [panels (e) and (f)]. For each material, two dispersion relation plots are shown. One panel displays the angular frequency $\omega$ as a function of the real part of the wave vector $k_z$, while the other panel shows $\omega$ as a function of the magnitude of $k_z$.

The modal analysis reveals that all membranes support corner and edge modes. Only modes with evanescent electromagnetic field characterized by penetration depth in vacuum $\delta_0 \gtrsim d$, where $\delta_0 \approx |k_{x0}|^{-1}$, contributes to NFRHT between the membranes. In the electrostatic limit far from the light line in vacuum, the magnitude of the $x$-component of the vacuum wave vector is approximated by $|k_{x0}| \approx |k_z|$, such that the penetration depth can be estimated from the $z$-component of the wave vector as $\delta_0 \approx |k_z|^{-1}$. The maximum contributing $z$-component of the wave vector dominating NFRHT [35] is therefore estimated as $|k_{z,\max}| \approx d^{-1}$ which corresponds to a value of $10^7$ rad/m for a gap spacing of 100 nm. This limit is identified by a vertical dashed line in all panels of Fig. 3.

Material losses, described by the imaginary part of the dielectric function, are negligibly small in the Reststrahlen band of SiC except near the transverse optical phonon frequency (see Fig. S1 in the Supplemental Material [33]) such that $\text{Re}(k_z) \gg \text{Im}(k_z)$. The expected resonant frequencies of the spectral heat transfer coefficient can therefore be determined from either Fig. 3(a) or 3(b), where $\omega$ is respectively plotted as a function of $\text{Re}(k_z)$ and $|k_z|$. Here, resonances are expected at frequencies of approximately $1.61 \times 10^{14}$ rad/s, $1.66 \times 10^{14}$ rad/s, and $1.73 \times 10^{14}$ rad/s for 20-nm, 40-nm, and 100-nm-thick membranes, respectively. These predictions are in good agreement with the resonances of the spectral heat transfer coefficient, $h_{\text{rad},\omega}$, calculated with the DSGF method and shown in Fig. 4(a). When compared to $h_{\text{rad},\omega}$ between infinite SiC surfaces which resonates at a frequency of $1.77 \times 10^{14}$ rad/s, corner and edge modes induce a resonance redshift as the membrane thickness decreases. In addition, spectral broadening is seen and is mediated by the large increase of the imaginary part of the dielectric function as the resonance is redshifted towards the transverse optical phonon frequency. The small losses in most of the Reststrahlen band of SiC, promoting strong SPhP coupling in the membranes, also explain why corner and edge modes have a noticeable impact on NFRHT in the thick membrane regime ($2 \lesssim t/d \lesssim 10$), where $G_{\text{rad}}$ follows a $t^{0.75}$ power law [Fig. 2(b)].

The $h_{\text{rad},\omega}$ resonance between SiN membranes redshifts and slightly broadens as $t$ decreases [Fig. 4(b)]. Here, however, the dispersion relations in terms of real wave vector cannot explain the resonance redshift. Indeed, Fig. 3(c) incorrectly suggests that all dispersion branches with resonance near $1.85 \times 10^{14}$ rad/s contribute to NFRHT independently of the membrane thickness. This is because material losses are non-negligible in the whole Reststrahlen band of SiN (see Fig. S2 in the Supplemental Material [33]), such that the dispersion relations in terms of $|k_z|$ should be used. Figure 3(d) suggests a $h_{\text{rad},\omega}$ resonance at a frequency slightly below $1.67 \times 10^{14}$ rad/s for 20-nm-thick membranes, and predicts $h_{\text{rad},\omega}$ resonances at frequencies of approximately $1.73 \times 10^{14}$ rad/s and $1.93 \times 10^{14}$ rad/s for the 40 nm and 100 nm cases, which is in good agreement with Fig. 4(b). Here, the enhancement of the heat transfer coefficient with respect to the infinite surface case [Fig. 2(a)] is not as large as for SiC for two reasons. First, resonance broadening with decreasing $t$ is modest compared to SiC. The imaginary part of the dielectric function in the Reststrahlen band of SiC increases by two orders of magnitude as the frequency decreases, whereas that of SiN increases by one order of magnitude. Second, the non-negligible losses in the entire Reststrahlen band of SiN reduce the maximum real wave vector that can be tunneled between the membranes. Note that $\text{Re}(k_z)$ is an indicator of the density of electromagnetic states [36], such that decreasing its value implies that less modes contribute to NFRHT. For instance, for 20-nm-thick membranes, the maximum real wave vector is limited to $\text{Re}(k_{z,\text{max}}) \approx 10^6$ rad/m at the resonant frequency of $\sim 1.67 \times 10^{14}$ rad/s [Fig. 3(c)]. For $t$ values of 40 nm and 100 nm, $\text{Re}(k_{z,\text{max}})$ increases to $\sim 2.3 \times 10^6$ rad/m and then decreases to $\sim 1.5 \times 10^6$ rad/m, which is consistent with the $h_{\text{rad}}$ trend reported in Fig. 2(a).

As in previous cases, resonances of the spectral heat transfer coefficient between $SiO_2$ membranes are redshifted compared to that between infinite surfaces [Fig. 4(c)]. Note that $SiO_2$ supports two Reststrahlen bands. Dispersion relations in the low-frequency Reststrahlen band are shown in Figs. 3(e) and 3(f) and analyzed hereafter, whereas the dispersion relations in the high-frequency Reststrahlen bands are presented in Sec. 3 of the Supplemental Material [33]. Material losses in the Reststrahlen bands of $SiO_2$ are non-negligible (see Fig. S3 of the Supplemental Material [33]), such that dispersion relations in terms of $|k_z|$ should be used to predict resonance redshift, as for SiN. According to Fig. 3(f), the low-frequency resonance of the spectral heat transfer coefficient is expected at a frequency slightly below $8.7 \times 10^{13}$ rad/s for 20-nm-thick membranes, whereas resonances are predicted at frequencies of approximately $8.75 \times 10^{13}$ rad/s and $9.30 \times 10^{13}$ rad/s for the 40 nm and 100 nm cases, which is in good agreement with Fig. 4(c). Here, material losses severely reduce the density of electromagnetic states. In the case of 20-nm-thick membranes, for example, the maximum real wave vector contributing to NFRHT is smaller than $10^6$ rad/m. This results in corner and edge mode-induced resonances that have lower magnitude than their infinite surface counterparts. Here, $\text{Re}(k_{z,\text{max}})$ monotonically decreases as $t$ is reduced, which is consistent with the $h_{\text{rad}}$ trend seen in Fig. 2(a). Combined with the lack of resonance spectral broadening, this explains the attenuation of the heat transfer coefficient with decreasing the membrane thickness shown in Fig. 2(a). The same conclusions hold for the resonance in the high-frequency Reststrahlen band.

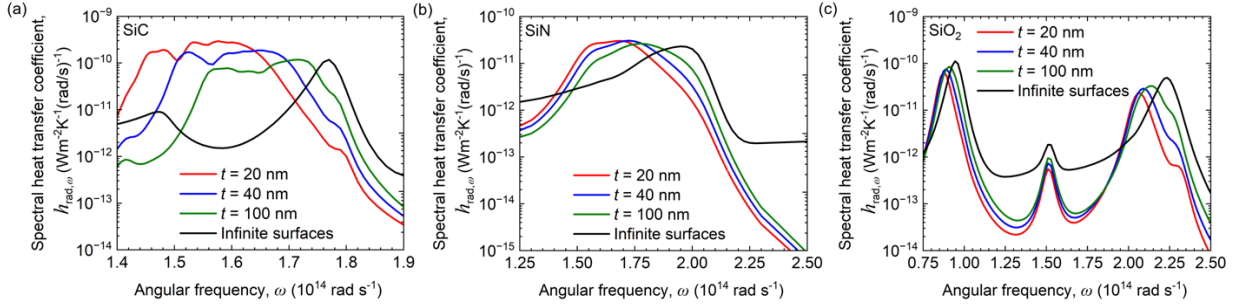

FIG 4. Spectral heat transfer coefficient, $h_{\text{rad},\omega}$, at 300 K for membrane thicknesses of 20 nm, 40 nm, and 100 nm calculated with the DSGF method. Results are compared against those obtained between infinite surfaces. (a) SiC, (b) SiN. (c) SiO$_2$.

In summary, DSGF simulations of NFRHT between SiC, SiN, and SiO$_2$ membranes combined with a modal analysis have shown that all membranes support electromagnetic corner and edge modes resulting in resonance redshifts of the spectral heat transfer coefficient. However, these corner and edge modes can induce large enhancement, mild enhancement and even attenuation of the heat transfer coefficient with respect to that between infinite surfaces. The NFRHT enhancement or attenuation in the dual nanoscale regime is directly related to material losses, described by the imaginary part of the dielectric function, which reduce the density of available electromagnetic states between the membranes. NFRHT is strongly enhanced with membranes made of polaritonic materials characterized by low losses in most of their Reststrahlen band, which promote SPhP coupling, combined with a large increase of their imaginary part of the dielectric function at the low-frequency edge of their Reststrahlen band, which induces substantial resonance spectral broadening. The results from this study will guide the design of novel contactless thermal management technologies.

## ACKNOWLEDGMENTS

This work was supported by the National Science Foundation (Grant No. CBET-1952210). Any opinions, findings, and conclusions or recommendations expressed in this material are those of the authors and do not necessarily reflect the views of the National Science Foundation. The support and resources from the Center for High Performance Computing at the University of Utah are gratefully acknowledged.

## END MATTER

*Appendix A: Calculation of the system Green's function* – The monochromatic system Green's function, $\bar{\bar{\mathbf{G}}}(\mathbf{r}, \mathbf{r}', \omega)$, relating the electric field from a point source $\mathbf{r}'$ to an observation point $\mathbf{r}$ is expressed in terms of the known free-space Green's function, $\bar{\bar{\mathbf{G}}}^0(\mathbf{r}, \mathbf{r}', \omega)$, via Dyson's equation [37]:

$$\bar{\bar{\mathbf{G}}}(\mathbf{r}, \mathbf{r}', \omega) = \bar{\bar{\mathbf{G}}}^0(\mathbf{r}, \mathbf{r}', \omega) + k_0^2 \int_{V_{\text{therm}}} \bar{\bar{\mathbf{G}}}^0(\mathbf{r}, \mathbf{r}'', \omega)[\varepsilon(\mathbf{r}'', \omega) - 1]\bar{\bar{\mathbf{G}}}(\mathbf{r}'', \mathbf{r}', \omega) \, d^3\mathbf{r}'', \tag{A1}$$

where $V_{\text{therm}}$ is the combined volume of the two membranes. Equation (A1) results in the following system of linear equations after discretizing the membranes into $N$ cubic subvolumes:

$$\left\{ \begin{bmatrix} \bar{\bar{\mathbf{I}}} & 0 & 0 \\ 0 & \ddots & 0 \\ 0 & 0 & \bar{\bar{\mathbf{I}}} \end{bmatrix} - k_0^2 \begin{bmatrix} \bar{\bar{\mathbf{G}}}^0_{11} & \cdots & \bar{\bar{\mathbf{G}}}^0_{1N} \\ \vdots & \ddots & \vdots \\ \bar{\bar{\mathbf{G}}}^0_{N1} & \cdots & \bar{\bar{\mathbf{G}}}^0_{NN} \end{bmatrix} \begin{bmatrix} \alpha_1^{(0)} & 0 & 0 \\ 0 & \ddots & 0 \\ 0 & 0 & \alpha_N^{(0)} \end{bmatrix} \right\} \begin{bmatrix} \bar{\bar{\mathbf{G}}}_{11} & \cdots & \bar{\bar{\mathbf{G}}}_{1N} \\ \vdots & \ddots & \vdots \\ \bar{\bar{\mathbf{G}}}_{N1} & \cdots & \bar{\bar{\mathbf{G}}}_{NN} \end{bmatrix}$$
$$= \begin{bmatrix} \bar{\bar{\mathbf{G}}}^0_{11} & \cdots & \bar{\bar{\mathbf{G}}}^0_{1N} \\ \vdots & \ddots & \vdots \\ \bar{\bar{\mathbf{G}}}^0_{N1} & \cdots & \bar{\bar{\mathbf{G}}}^0_{NN} \end{bmatrix}, \tag{A2}$$

where $\alpha_i^{(0)} = \Delta V_i[\varepsilon(\mathbf{r}_i, \omega) - 1]$ is the bare polarizability. The discretized free-space system Green's function in vacuum is calculated as:

$$\bar{\bar{\mathbf{G}}}^0(\mathbf{r}_i, \mathbf{r}_j, \omega) = \frac{\exp(ik_0 r_{ij})}{4\pi r_{ij}} \left[ \begin{array}{c} \left(1 - \frac{1}{(k_0 r_{ij})^2} + \frac{i}{k_0 r_{ij}}\right) \bar{\bar{\mathbf{I}}} \\ -\left(1 - \frac{3}{(k_0 r_{ij})^2} + \frac{3i}{k_0 r_{ij}}\right) (\hat{\mathbf{r}}_{ij} \hat{\mathbf{r}}_{ij}^\dagger) \end{array} \right], j \neq i, \tag{A3}$$

where lattice locations $\mathbf{r}_i$ and $\mathbf{r}_j$ are represented by subscripts $i$ and $j$, $r_{ij} = |\mathbf{r}_i - \mathbf{r}_j|$, and $\hat{\mathbf{r}}_{ij} = \frac{(\mathbf{r}_i - \mathbf{r}_j)}{|\mathbf{r}_i - \mathbf{r}_j|}$. At the singularity point where $\mathbf{r}_i = \mathbf{r}_j$, the discretized free-space system Green's function derived from the principal value method is given by [38]:

$$\bar{\bar{\mathbf{G}}}^0(\mathbf{r}_i, \mathbf{r}_j, \omega) = \frac{\bar{\bar{\mathbf{I}}}}{3\Delta V_j k_0^2} \{2[e^{ia_j k_0}(1 - ia_j k_0) - 1] - 1\}, j = i \tag{A4}$$

19where $a_j = (3\Delta V_j/4\pi)^{1/3}$.

The thermal conductance and heat transfer coefficient between the membranes are calculated from the system Green's function obtained from the solution of Eq. (A2).

*Appendix B: NFRHT between two infinite surfaces* – The heat transfer coefficient between two infinite surfaces is calculated using a closed-form expression derived from fluctuational electrodynamics [34]:

$$h_{\text{rad,inf}}(T) = \frac{1}{2\pi}\int_0^\infty d\omega \left.\frac{\partial\Theta(\omega,T')}{\partial T}\right|_{T'=T}$$

$$\times \left[\begin{array}{c}\frac{1}{2\pi}\int_0^{k_0} dk_\rho k_\rho \sum_{\gamma=TE,TM} \frac{\left(1-|r_{0h}^\gamma|^2\right)\left(1-|r_{0c}^\gamma|^2\right)}{|1-r_{0h}^\gamma r_{0c}^\gamma e^{2i\text{Re}(k_{x0})d}|^2} \\ +\frac{2}{\pi}\int_{k_0}^\infty dk_\rho k_\rho e^{-2\text{Im}(k_{x0})d}\sum_{\gamma=TE,TM}\frac{\text{Im}(r_{0h}^\gamma)\text{Im}(r_{0c}^\gamma)}{|1-r_{0h}^\gamma r_{0c}^\gamma e^{-2\text{Im}(k_{x0})d}|^2}\end{array}\right] \quad (B1)$$

where $k_\rho$ is the wave vector parallel to the *y-z* plane, $k_{x0}$ is the *x*-component of the vacuum wave vector normal to the surfaces, and $r_{0h}^\gamma$ and $r_{0c}^\gamma$ are the $\gamma$-polarized Fresnel reflection coefficients at the vacuum-hot (0-*h*) and vacuum-cold (0-*c*) interfaces, respectively.

When material losses are significant, the resonance of the spectral heat transfer coefficient between infinite surfaces should be estimated by evaluating the frequency at which $|\varepsilon + 1|$ is minimum [39]. Resonances are predicted at frequencies of $1.77\times10^{14}$ rad/s for SiC, $2.03\times10^{14}$ rad/s for SiN, and $9.67\times10^{13}$ rad/s and $2.24\times10^{14}$ rad/s for SiO$_2$. These predictions are in good agreement with the spectral heat transfer presented in Fig. 4. Note that for SiC characterized by a small imaginary part of the dielectric function in most of its Reststrahlen band, resonance can be predicted by determining the frequency at which $\text{Re}(\varepsilon) = -1$ [4].

# Supplemental Material

# Near-field radiative heat transfer in the dual nanoscale regime between polaritonic membranes

Lívia Corrêa McCormack[1], Lei Tang[2] and Mathieu Francoeur[1*]

Department of Mechanical Engineering, The University of Utah, Salt Lake City, UT 84112, USA

[2]Department of Mechanical Engineering, University of California at Berkeley, Berkeley, CA 94720, USA

*Corresponding author: mfrancoeur@mech.utah.edu

# S1. DIELECTRIC FUNCTIONS

## A. SiC

The dielectric function of polycrystalline SiC is described by a Lorentz model [1]:

$$\varepsilon(\omega) = \varepsilon_\infty \frac{\omega^2 - \omega_{LO}^2 + i\Gamma\omega}{\omega^2 - \omega_{TO}^2 + i\Gamma\omega} \tag{S1}$$

where $\varepsilon_\infty = 8$, $\omega_{LO} = 1.801 \times 10^{14}$ rad/s, $\omega_{TO} = 1.486 \times 10^{14}$ rad/s, and $\Gamma = 3.767 \times 10^{12}$ 1/s are respectively the high-frequency dielectric constant, the longitudinal optical phonon frequency, the transverse optical phonon frequency, and the damping constant. The real and imaginary parts of the dielectric function of SiC are shown in Fig. S1.

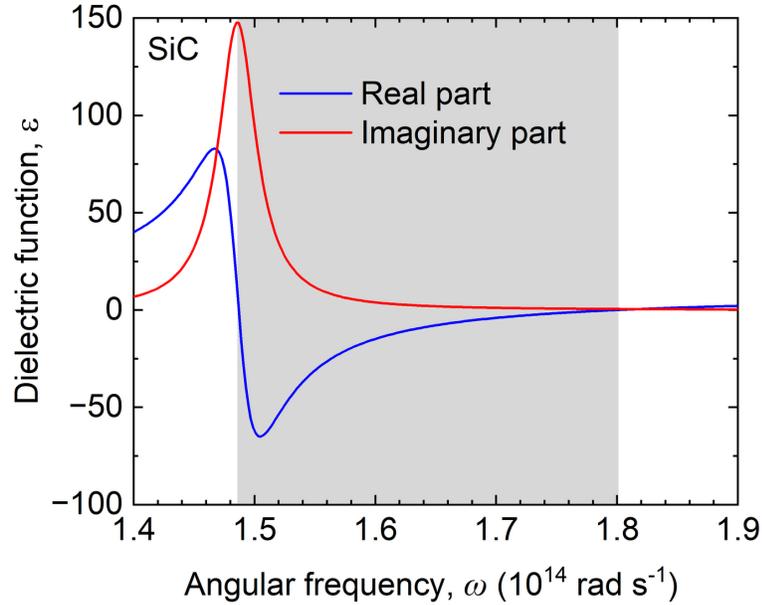

FIG S1. Real and imaginary parts of the dielectric function of SiC. The Reststrahlen band is identified by the gray-shaded box.

## B. SiN

The dielectric function of SiN is described by the Maxwell-Helmholtz-Drude dispersion model [2]:

$$\varepsilon(\omega) = \varepsilon_\infty + \sum_{k=1}^{M} \frac{\Delta\varepsilon_k \omega_{T_k}^2}{\omega_{T_k}^2 - \omega^2 - i\omega\Gamma_k \exp\left[-\alpha_k\left(\frac{\omega_{T_k}^2 - \omega^2}{\omega\Gamma_k}\right)^2\right]} \quad (S2)$$

where $\varepsilon_\infty$ is the high-frequency dielectric constant, $M$ is the number of oscillators, $\Delta\varepsilon_k$ is the difference in dielectric function between adjacent oscillators, $\omega_{T_k}$ is the oscillator resonance frequency, $\Gamma_k$ is the damping coefficient, and $\alpha_k$ is an interpolation constant. The numerical parameters needed to calculate Eq. (S2) are provided in Table 1 of Ref. [2]. The real and imaginary parts of the dielectric function of SiN are shown in Fig. S2.

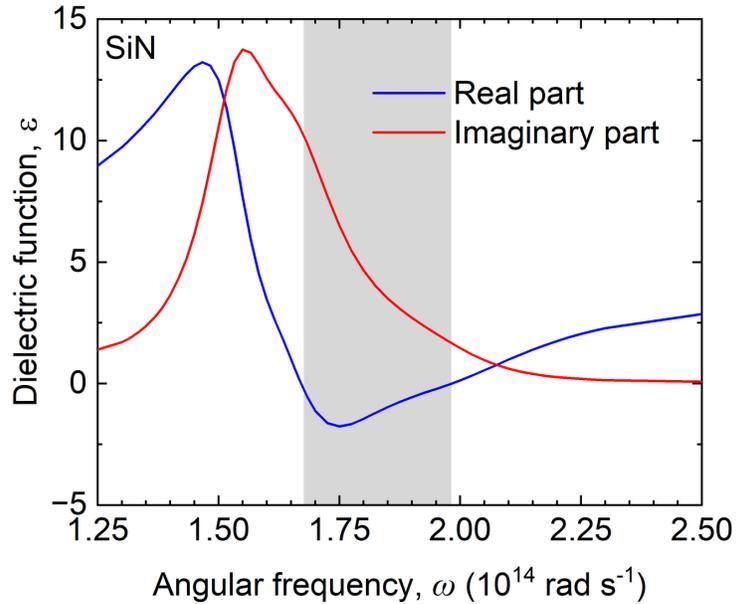

FIG S2. Real and imaginary parts of the dielectric function of SiN. The Reststrahlen band is identified by the gray-shaded box.

## C. SiO$_2$

The dielectric function of SiO$_2$ is described by the following Lorentz oscillator model [3]:

$$\varepsilon(\omega) = \varepsilon_\infty + \sum_{k=1}^{M} \frac{S_k}{1-\left(\frac{\omega}{\omega_{0,k}}\right)^2 - i\Gamma_k\left(\frac{\omega}{\omega_{0,k}}\right)} \tag{S3}$$

where $\varepsilon_\infty$ is the high-frequency dielectric constant, $M$ is the number of oscillators, $S_k$ is a fitting parameter, $\omega_{0,k}$ is the oscillator resonance frequency, and $\Gamma_k$ is the damping coefficient. The numerical parameters needed to calculate Eq. (S3) are provided in Table 3 of Ref. [3]. The real and imaginary parts of the dielectric function of SiO$_2$ are shown in Fig. S3.

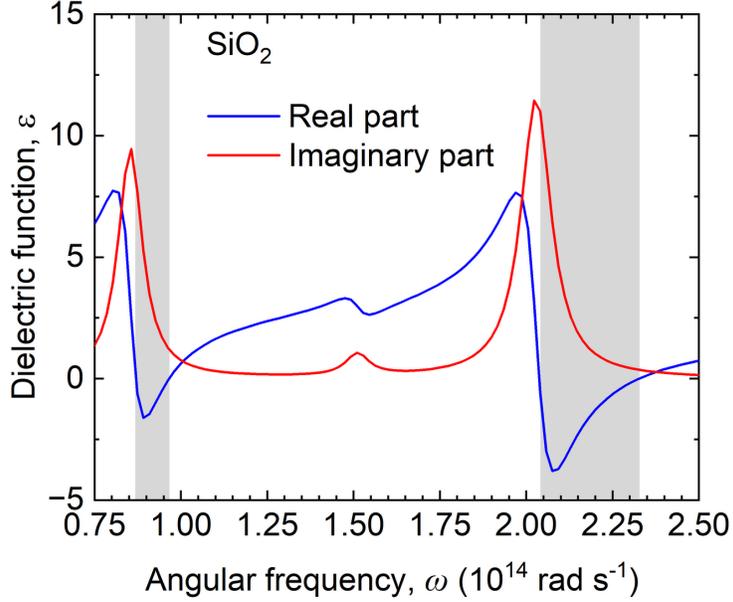

FIG S3. Real and imaginary parts of the dielectric function of SiO$_2$. The Reststrahlen bands are identified by the gray-shaded boxes.

## S2. CONVERGENCE ANALYSIS OF THE DSGF

As an example, the convergence of the DSGF is discussed hereafter for 100-nm-thick SiC membranes separated by a vacuum gap $d = 100$ nm. The same convergence analysis was performed for all materials and membrane thicknesses.

Figure S4 shows the spectral conductance for three discretization schemes: $N = 1{,}600$ uniform subvolumes, $N = 12{,}800$ uniform subvolumes, and $N = 25{,}000$ uniform subvolumes, which corresponds to subvolume side lengths $L_{sub}$ of respectively 50 nm, 25 nm, and 20 nm.

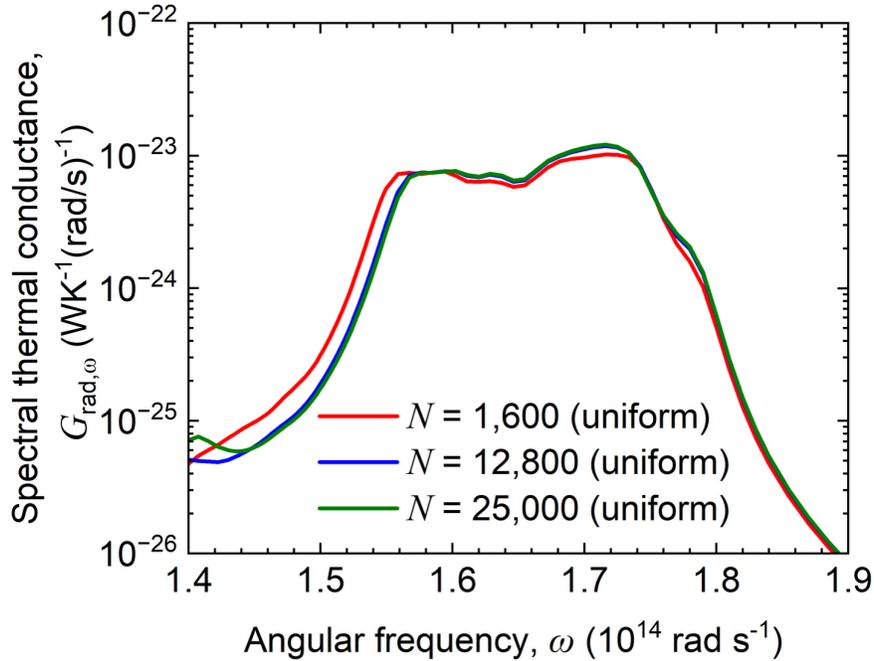

FIG S4. Spectral thermal conductance at 300 K between two 100-nm-thick SiC membranes separated by $d = 100$ nm. Results are shown for three discretization schemes: $N = 1{,}600$ uniform subvolumes ($L_{sub} = 50$ nm), $N = 12{,}800$ uniform subvolumes ($L_{sub} = 25$ nm), and $N = 25{,}000$ uniform subvolumes ($L_{sub} = 20$ nm).

Figure S4 indicates that the spectral conductance is nearly identical for 12,800 and 25,000 uniform subvolumes. This is further confirmed by comparing the total (i.e., spectrally integrated) conductance. The relative difference of the total conductance when increasing the number of subvolumes from 12,800 and 25,000 is ~0.5%. As such, it is concluded that convergence is reached with 12,800 uniform subvolumes of side length $L_{sub}$ = 25 nm.

Minimization of the computational cost associated with the DSGF is achieved by using nonuniform discretization, since ~70% of the total power is dissipated in the cold membrane within the first 200 nm measured along the $x$-direction with respect to the gap spacing (see Fig. 1 in the main manuscript). Three nonuniform discretization schemes are tested:

1) $N$ = 2,720 subvolumes, $L_{sub}$ = 25 nm for $x$ = 0 to 100 nm, $L_{sub}$ = 50 nm for $x$ = 100 to 1000 nm;

2) $N$ = 3,840 subvolumes, $L_{sub}$ = 25 nm for $x$ = 0 to 200 nm, $L_{sub}$ = 50 nm for $x$ = 200 to 1000 nm;

3) $N$ = 4,960 subvolumes, $L_{sub}$ = 25 nm for $x$ = 0 to 300 nm, $L_{sub}$ = 50 nm for $x$ = 300 to 1000 nm

where $x$ is defined with respect to the vacuum gap spacing. Figure S5 compares the spectral conductance resulting from the three nonuniform discretization schemes described above against that obtained with the optimal uniform discretization ($N$ = 12,800, $L_{sub}$ =25 nm).

Figure S5 indicates that the spectral conductance with 4,960 nonuniform subvolumes is nearly identical to that with 12,800 subvolumes. This is further confirmed by comparing the total conductance. The relative difference of the total conductance is ~0.8%. As such, it is concluded that convergence is reached with 4,960 nonuniform subvolumes.

All DSGF simulations presented in the main manuscript have been performed using nonuniform discretizaton. The number of nonuniform subvolumes ensuring converged DSGF results for SiC as a function of the membrane thickness are shown in Table S1. The same information is provided in Table S2 for SiN and $SiO_2$ (the same discretization scheme was used for both materials).

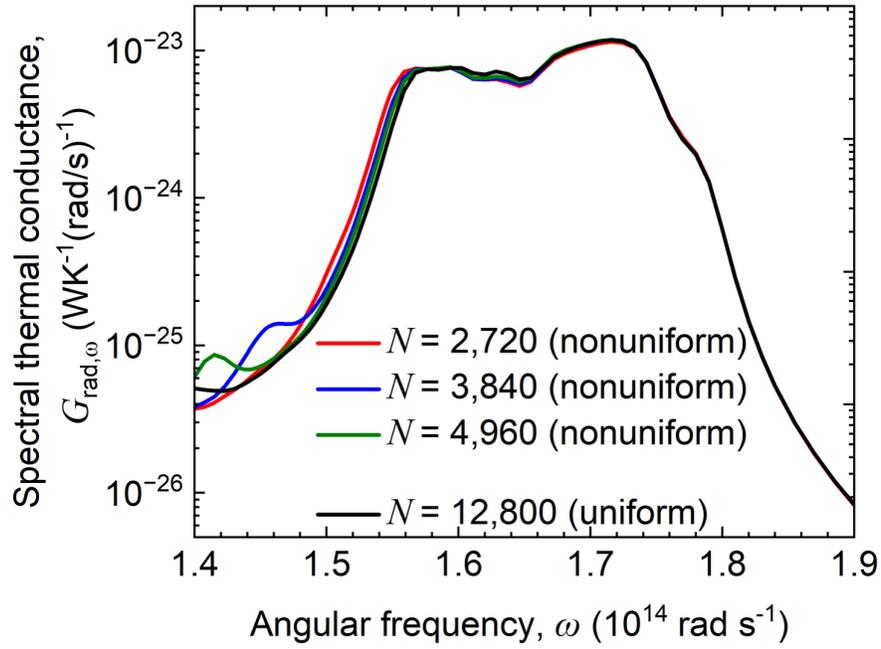

FIG S5. Spectral thermal conductance at 300 K between two 100-nm-thick SiC membranes separated by $d = 100$ nm. Results are shown for $N = 2{,}720$ nonuniform subvolumes, $N = 3{,}840$ nonuniform subvolumes, $N = 4{,}960$ nonuniform subvolumes, and $N = 12{,}800$ uniform subvolumes.

TABLE S1. Number of nonuniform subvolumes for converged DSGF results for SiC as a function of the membrane thickness. The distance $x$ is defined with respect to the vacuum gap spacing.

| Membrane thickness $t$ | Total number of nonuniform subvolumes $N$ | Subvolume side length $L_{sub}$ |
|---|---|---|
| 20 nm | 17,600 | 5 nm ($x$ = 0 to 20 nm)<br>10 nm ($x$ = 20 to 200 nm)<br>20 nm ($x$ = 200 to 1000 nm) |
| 40 nm | 14,900 | 10 nm ($x$ = 0 to 120 nm)<br>20 nm ($x$ = 120 to 600 nm)<br>40 nm ($x$ = 600 to 1000 nm) |
| 60 nm | 10,824 | 15 nm ($x$ = 0 to 210 nm)<br>30 nm ($x$ = 210 to 1000 nm) |
| 80 nm | 11,200 | 16 nm ($x$ = 0 to 240 nm)<br>40 nm ($x$ = 240 to 1000 nm) |
| 100 nm | 4,960 | 25 nm ($x$ = 0 to 300 nm)<br>50 nm ($x$ = 300 to 1000 nm) |
| 200 nm | 10,320 | 20 nm ($x$ = 0 to 200 nm)<br>100 nm ($x$ = 200 to 1000 nm) |
| 400 nm | 10,160 | 20 nm ($x$ = 0 to 100 nm)<br>100 nm ($x$ = 100 to 200 nm)<br>200 nm ($x$ = 200 to 1000 nm) |
| 600 nm | 7,920 | 25 nm ($x$ = 0 to 100 nm)<br>100 nm ($x$ = 100 to 200 nm)<br>200 nm ($x$ = 200 to 1000 nm) |
| 800 nm | 10,560 | 25 nm ($x$ = 0 to 100 nm)<br>100 nm ($x$ = 100 to 200 nm)<br>200 nm ($x$ = 200 to 1000 nm) |
| 1000 nm | 13,608 | 25 nm ($x$ = 0 to 100 nm)<br>100 nm ($x$ = 100 to 500 nm)<br>200 nm ($x$ = 500 to 1000 nm) |

TABLE S2. Number of nonuniform subvolumes for converged DSGF results for SiN and SiO$_2$ as a function of the membrane thickness. The distance $x$ is defined with respect to the vacuum gap spacing.

| Membrane thickness $t$ | Total number of nonuniform subvolumes $N$ | Subvolume side length $L_{sub}$ |
|---|---|---|
| 20 nm | 12,000 | 10 nm ($x$ = 0 to 200 nm)<br>20 nm ($x$ = 200 to 1000 nm) |
| 40 nm | 14,900 | 10 nm ($x$ = 0 to 120 nm)<br>20 nm ($x$ = 120 to 600 nm)<br>40 nm ($x$ = 600 to 1000 nm) |
| 60 nm | 6,732 | 20 nm ($x$ = 0 to 220 nm)<br>30 nm ($x$ = 220 to 1000 nm) |
| 80 nm | 6,700 | 20 nm ($x$ = 0 to 240 nm)<br>40 nm ($x$ = 240 to 1000 nm) |
| 100 nm | 4,960 | 25 nm ($x$ = 0 to 300 nm)<br>50 nm ($x$ = 300 to 1000 nm) |
| 200 nm | 5,440 | 25 nm ($x$ = 0 to 200 nm)<br>100 nm ($x$ = 200 to 1000 nm) |
| 400 nm | 5,280 | 25 nm ($x$ = 0 to 100 nm)<br>100 nm ($x$ = 100 to 200 nm)<br>200 nm ($x$ = 200 to 1000 nm) |
| 600 nm | 7,920 | 25 nm ($x$ = 0 to 100 nm)<br>100 nm ($x$ = 100 to 200 nm)<br>200 nm ($x$ = 200 to 1000 nm) |
| 800 nm | 10,560 | 25 nm ($x$ = 0 to 100 nm)<br>100 nm ($x$ = 100 to 200 nm)<br>200 nm ($x$ = 200 to 1000 nm) |
| 1000 nm | 13,608 | 25 nm ($x$ = 0 to 100 nm)<br>100 nm ($x$ = 100 to 500 nm)<br>200 nm ($x$ = 500 to 1000 nm) |

## S3. CORNER AND EDGE MODE DISPERSION RELATIONS IN THE HIGH-FREQUENCY RESTSTRAHLEN BAND OF $SiO_2$

Corner and edge mode dispersion relations in the high-frequency Reststrahlen band of $SiO_2$ is shown in Fig. S6. As for the low-frequency resonance discussed in the main text, material losses in the high-frequency Reststrahlen band of $SiO_2$ reduce the maximum real wave vector values, resulting in a drop of the available density of electromagnetic states. This explains the lower magnitude of the heat transfer coefficient between membranes compared to that between infinite surfaces.

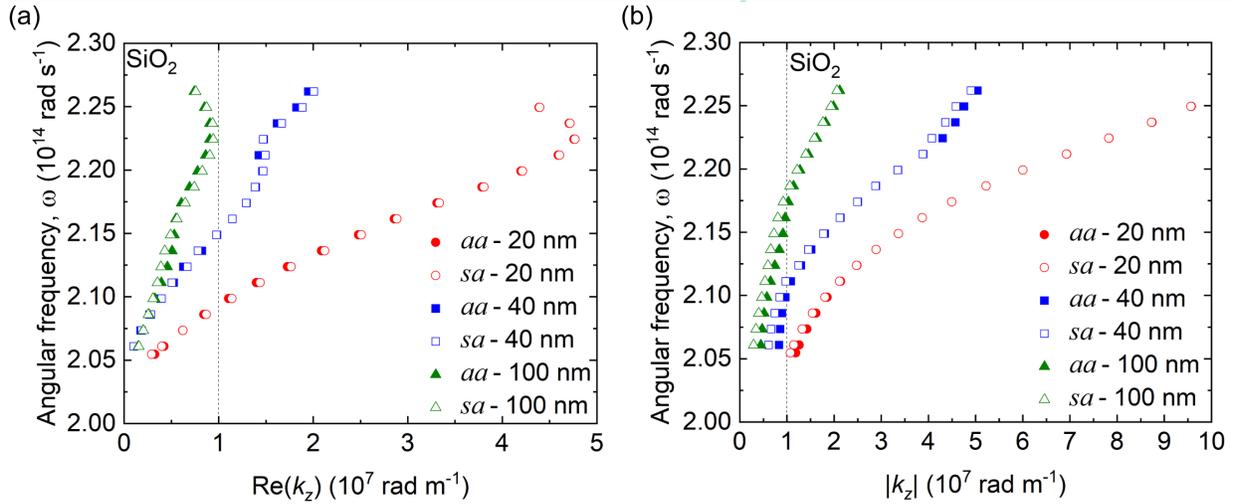

FIG S6. Dispersion relations in the high-frequency Reststrahlen band of $SiO_2$ for the fundamental corner and edge modes, *aa* and *sa*, in a solitary membrane of infinite length $L$ and varying thickness $t$ calculated via COMSOL Multiphysics. Panel (a) displays the angular frequency $\omega$ as a function of the real part of the wave vector $k_z$, while panel (b) shows $\omega$ as a function of the magnitude of $k_z$.